\documentclass[aps,pre,twocolumn,superscriptaddress, longbibliography, 10pt]{revtex4-2}
\usepackage{amsmath,amssymb,graphicx,bm}
\usepackage[x11names]{xcolor}
\usepackage[colorlinks=true,linkcolor=blue,citecolor=blue,urlcolor=blue]{hyperref}
\usepackage{soul }
\usepackage{siunitx}
\usepackage{placeins}

\newcommand{\rev}[1]{\textcolor{red}{#1}}
\renewcommand{\rev}[1]{#1}
\newcommand{\rrev}[1]{\textcolor{red}{#1}}
\renewcommand{\rrev}[1]{#1}

\DeclareSIUnit{\molar}{M}
\newcommand{\tr}{\text{Tr}}
\newcommand{\var}{\operatorname{Var}}
\usepackage[normalem]{ulem}

\begin{document}

\title{Fully developed active turbulence defined through a non-equilibrium
phase transition}
\author{Lasse Bonn}
\thanks{These authors contributed equally.}
\affiliation{Niels Bohr Institute, University of Copenhagen, Blegdamsvej 17, Copenhagen 2100, Denmark}
\author{Tianxiang Ma}
\thanks{These authors contributed equally.}
\affiliation{Niels Bohr Institute, University of Copenhagen, Blegdamsvej 17, Copenhagen 2100, Denmark}
\author{Olga Bantysh}
\affiliation{Department of Materials Science and Physical Chemistry, Universitat de Barcelona, 08028 Barcelona, Spain}
\affiliation{Institute of Nanoscience and Nanotechnology, IN2UB, Universitat de Barcelona, 08028 Barcelona, Spain}
\author{Wei Feng}
\affiliation{School of Physics, and Peng Huanwu Center for Fundamental Theory, Shaanxi Key Laboratory for Theoretical Physics Frontiers, Northwest University, 710127, Xi'An, China}
\author{Martin Cramer Pedersen}
\affiliation{Niels Bohr Institute, University of Copenhagen, Blegdamsvej 17, Copenhagen 2100, Denmark}
\author{Guangyin Jing}
\email{Corresponding author, email: jing@nwu.edu.cn}
\affiliation{School of Physics, and Peng Huanwu Center for Fundamental Theory, Shaanxi Key Laboratory for Theoretical Physics Frontiers, Northwest University, 710127, Xi'An, China}
\author{Jordi Ign\'es-Mullol}
\affiliation{Department of Materials Science and Physical Chemistry, Universitat de Barcelona, 08028 Barcelona, Spain}
 \affiliation{Institute of Nanoscience and Nanotechnology, IN2UB, Universitat de Barcelona, 08028 Barcelona, Spain}
 \author{Francesc Sagu\'es}
\affiliation{Department of Materials Science and Physical Chemistry, Universitat de Barcelona, 08028 Barcelona, Spain}
 \affiliation{Institute of Nanoscience and Nanotechnology, IN2UB, Universitat de Barcelona, 08028 Barcelona, Spain}
\author{Nuno A. M. Araujo}
\affiliation{Departamento de Física, Faculdade de Ciências, Universidade de Lisboa, Lisboa, Portugal}
\affiliation{Centro de Física Teórica e Computacional, Faculdade de Ciências, Universidade de Lisboa, Lisboa, Portugal}
\author{Amin Doostmohammadi}
\email{Corresponding author, email: doostmohammadi@nbi.ku.dk}
\affiliation{Niels Bohr Institute, University of Copenhagen, Blegdamsvej 17, Copenhagen 2100, Denmark}

\date{28. September 2026}

\begin{abstract}
Non-equilibrium systems challenge the standard definitions of phases and phase transitions from equilibrium statistical physics. For example, in active fluids continuous energy injection at the microscale drives the system intrinsically out of equilibrium. This process gives rise to collective turbulent-like flows whose onset remains poorly defined. Here we show that the onset corresponds to an activity-driven phase transition, marked by the emergence of a system-spanning critical backbone of vorticity nodal lines. Combining two independent experiments, microtubule kinesin active nematics and dense suspensions of swimming bacteria, with large-scale simulations of active nematics and fluctuating nematohydrodynamics, we show that this transition occurs at a critical activity threshold. Below this threshold, vorticity structures are fragmented and vortex centres form only finite, mechanically floppy networks. Above this threshold, the nodal lines of the vorticity field percolate and the vortex centres form a rigid, system-spanning cluster. The geometric transition is characterised by the emergence of critical percolation statistics for the vorticity nodal lines, while the mechanical transition reflects the emergence of a vortex configuration that shows rigidity percolation statistics for vortex centres. We show that both transitions are absent in equilibrium systems that obey detailed balance, demonstrating that the transition is intrinsically non-equilibrium. These results establish an experimentally accessible definition of \textit{fully developed active turbulence}, linking microscopic activity to macroscopic geometry, and mechanics of living systems.
\end{abstract}

\maketitle

Non-equilibrium systems challenge the conceptual foundations of phase behaviour developed within equilibrium statistical physics.
In equilibrium, phases and phase transitions are classified through free energy landscapes, symmetry breaking, and universal scaling laws, rooted in detailed balance and variational principles~\cite{gross1996role,falkovich_symmetries_2009}.
Driven systems, by contrast, operate under sustained energy injection and dissipation, often lacking detailed balance and even an underlying free energy~\cite{marchetti2013hydrodynamics,bowick2022symmetry}.
Whether such systems admit well defined phase transitions with universal properties, and how these transitions should be identified, remains a central open question across statistical physics, soft matter, and biological physics~\cite{alert2020universal,mcgreevy2023generalized}.

Active fluids form a particularly rich class of non-equilibrium matter, in which  microscopic constituents convert chemical energy into motion, generating collective flows that span many length scales.
Examples range from cytoskeletal assemblies driven by molecular motors and epithelial tissues to dense bacterial suspensions and cell monolayers~\cite{sanchez_spontaneous_2012, wensink_meso-scale_2012,duclos_spontaneous_2018,ventura2022multiciliated}.
At sufficiently strong driving, these systems develop chaotic, vortex dominated dynamics commonly referred to as active turbulence.
Despite extensive experimental and theoretical efforts, however, there is still no consensus on how to define the onset of fully developed active turbulence in a precise and quantitative way.
Unlike inertial turbulence, where asymptotic regimes and control parameters are well established, active turbulence lacks a clear criterion separating weakly disordered flows from a genuinely collective, system-scale state.

A promising direction for identifying such transitions lies in geometric and statistical descriptions of complex flows.
In two dimensional systems, the organisation of vorticity encodes information beyond conventional correlation functions, revealing large scale structures that can persist despite strong local fluctuations~\cite{bernard_conformal_2006}.
In equilibrium critical phenomena, percolation theory provides a paradigmatic example of how global connectivity emerges at a threshold, independent of microscopic details~\cite{stauffer2018introduction}.
Whether analogous geometric transitions occur in driven fluids and whether they signal genuine non-equilibrium phase transitions remains largely unexplored.

\rev{
A powerful framework to characterise such geometric transitions is provided by the theory of Schramm Loewner Evolution (SLE), a major development in mathematical physics that enables a rigorous classification of universality in two dimensional stochastic systems~\cite{schramm2000scaling, gruzberg_loewner_2004, cardy_sle_2005, bauer20062d}.
SLE defines a one parameter family of random conformally invariant curves, controlled by a diffusivity $\kappa$, which arise as the scaling limits of interfaces in many critical models of statistical physics.
Remarkably, different microscopic systems that share the same value of $\kappa$ generate, in the continuum limit, identical ensembles of curves and therefore belong to the same universality class.
The construction of these curves is based on the Loewner equation, which describes the evolution in a parameter $t$ of a conformal map.
This evolution is driven by a Brownian motion with variance set by $\kappa$, and the resulting trace defines a random curve whose statistical properties are fully determined by this single parameter.
Specific values of $\kappa$ have been rigorously linked to well known models: for example, loop erased random walks converge to $\kappa = 2$~\cite{lawler2011conformal}, interfaces in the critical Ising model correspond to $\kappa = 3$~\cite{chelkak_convergence_2014}, and cluster boundaries in critical percolation are described by $\kappa = 6$~\cite{smirnov2001critical}.
This correspondence provides a direct route to identifying universality classes in complex systems.
By testing whether the interfaces observed in a given system are consistent with SLE, and by measuring the associated value of $\kappa$, one can determine whether they share the same large scale statistics as established critical models.
Moreover, demonstrating SLE behaviour implies conformal invariance, that is invariance under translations, rotations, rescalings, and more general angle preserving transformations.
Among continuous symmetries, conformal symmetry is particularly restrictive, strongly constraining the geometry of fluctuations and fixing many statistical properties independently of microscopic details~\cite{francesco2012conformal}.
}

Recent experimental studies of active living matter have added a surprising element to this picture.
In systems as diverse as epithelial cell layers and bacterial colonies, the nodal lines of the vorticity field, zeros of the field that separate regions of opposite rotation, were found to exhibit statistical properties associated with critical percolation, that is, shown to be $SLE_6$~\cite{andersen_evidence_2025,ma2026cell}.
These observations suggest that active flows may self organise into states governed by universal geometric statistics, despite being far from equilibrium. 
At the same time, the generality, origin, and physical significance of this behaviour remain unclear.
In particular, it is not known whether such statistics reflect a robust transition controlled by activity or whether they are incidental features of specific biological systems.

An equally important and largely unexplored question concerns the mechanical organisation of active flows.
Vortices are not merely geometric objects but carriers of stress and momentum. 
Their interactions and organisation are therefore central to understanding the behaviour of the fluid.
In disordered solids, rigidity percolation describes the emergence of mechanical stability through the formation of a system spanning rigid network~\cite{jacobs1995generic}.
In biological systems, rigidity percolation has been observed at the intracellular level, playing a role in cell motility~\cite{garcia-arcos_rigidity_2024}, and at the tissue level regulating morphogenesis~\cite{petridou2021rigidity}.
Whether active turbulence undergoes an analogous mechanical transition, and how such a transition might relate to geometric connectivity in the vorticity field, has not been addressed.

In this work, we show that active fluids undergo an activity-driven phase transition characterised by the simultaneous emergence of geometric and mechanical connectivity in the flow.
Using two independent experimental systems, microtubule kinesin active nematics and dense suspensions of swimming bacteria, together with large scale simulations of active nematic and non-equilibrium fluctuating hydrodynamic models, we demonstrate that above a critical activity threshold the nodal lines of the vorticity field form a system spanning backbone with the statistics of critical percolation.
At the same threshold, vortex centres undergo rigidity percolation, assembling into a mechanically constrained network that spans the system.
Below this threshold, vorticity structures remain fragmented and vortex networks are mechanically floppy.
The transition is absent in equilibrium systems that respect detailed balance, establishing it as an intrinsically non-equilibrium phenomenon.

\begin{figure}[t]
\centering
\includegraphics[width=1\linewidth]{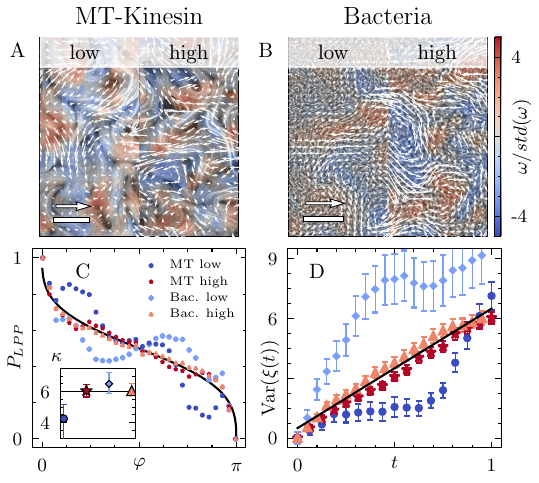}
\caption{\textbf{Emergence and breakdown of critical percolation statistics in experimental systems}
\textbf{(A-B)} Velocity and vorticity field normalised by standard deviation ($\omega/std(\omega)$) overlaid on 
\textbf{(A)} microtubule–kinesin (MT-Kinesin) active nematics (fluorescence image) under low (left) and high (right) activity, controlled by ATP concentration, and 
\textbf{(B)} dense bacterial suspensions (bright field image) under low (left) and high (right) activity, controlled by oxygen availability.
\rev{For clarity, only part of the experimental field of view is shown.}
\textbf{(C)} Left passage probability measurements at low (blues) and high (reds) activities.
High activity systems follow the black line which shows the analytical prediction for $\kappa=6$.
(Inset) Diffusivity $\kappa$ obtained from the left passage probability of nodal lines, showing that only the high activity state is consistent with the $SLE_{6}$ prediction $\kappa\simeq 6$\rrev{, with errors as defined in equation \ref{eq:lpperror}.}
\textbf{(D)} Driving function variance at low (blues) and high (reds) activities.
High activity systems follow the black line which indicates a slope of $\var(\xi(t)) = 6t$ and therefore $\kappa=6$.
\rrev{Data as mean $\pm$ S.E.M.}
\rrev{For (C) and (D), the number of frames analysed is $N_{MT\, low} = 29, N_{MT\, high} = 174, N_{Bac.\,low} = 50, N_{Bac.\,high}=105$.}
Scale bars: MT-kinesin: 
\rrev{\SI{0.5}{\micro\metre/\second} (white arrow)}, 
\SI{100}{\micro\metre} (white line);
Bacteria:
\SI{100}{\micro\metre/\second}, 
\SI{20}{\micro\metre}.
}
\label{fig:exp}
\end{figure}

\textit{Activity-driven emergence and breaking of critical percolation statistics.}—
 Originally developed for equilibrium critical phenomena, SLE has since uncovered conformal invariance in far broader contexts, from two-dimensional turbulence and wave chaos to rigidity percolation and liquid-crystal coarsening~\cite{bernard_conformal_2006,puggioni_conformal_2020, noseda2024conformal, javerzat_schramm-loewner_2024, almeida_critical_2025}. 
 The framework of Schramm–Loewner Evolution therefore offers a stringent and quantitative test not only for the presence of conformal symmetry, but also for determining the universality class of complex systems.
 To assess for critical percolation statistics, we focus on the {\it left passage probability} and the {\it Loewner driving function} $\xi(t)$, two independent tests \rev{for measuring $\kappa$} in the Schramm–Loewner framework (more details in \textbf{Methods}, section \ref{sec:SLESI} and \textbf{Fig.\ref{fig:ED1}}).
The left passage probability measures the likelihood that a curve passes to the left of a point $z=\rho e^{i\varphi}$ in the upper half-plane, where $\rho$ and $\varphi$ are the radius and polar angle \rev{(see \textbf{Fig.\ref{fig:ED2}A} for schematic)}. 
For a $SLE_{\kappa}$ trace, this probability has the exact form~\cite{schramm_percolation_2001}
\begin{equation}
\begin{aligned}
P_{LPP}(\varphi)=\tfrac{1}{2}
&+\frac{\Gamma(4/\kappa)}{\sqrt{\pi}\,\Gamma\!\left((8-\kappa)/(2\kappa)\right)} \\
&\cot(\varphi)\,{}_2F_1\!\left(\tfrac{1}{2},\tfrac{4}{\kappa};\tfrac{3}{2};-\cot^{2}\varphi\right),
\end{aligned}
\label{eq:lpp}
\end{equation}
where $\Gamma$ is Euler's gamma function and ${}_2F_1$ is the Gaussian hypergeometric function.
The only free parameter in this expression is the diffusivity $\kappa$ that we are set to determine~\cite{pose_shortest_2014}. For $\kappa=6$, Eq.~\eqref{eq:lpp} is Cardy's formula for critical percolation~\cite{cardy_conformal_2003}.

Independently, the driving function, obtained by inverting the Loewner mapping, captures the stochastic process that generates the contour: for conformally invariant curves, this function must perform a one-dimensional Brownian motion with variance $\langle \xi^{2}(t)\rangle=\kappa t$ and Gaussian statistics at every time point, where $t$ is the Loewner time that parametrises the growth of the curve (see details in \textbf{Methods}, section \ref{sec:SLESI} and \textbf{Fig.\ref{fig:ED2}C} for schematic).
These two measures offer stringent and independent tests for both conformal symmetry and the associated universality class.

We analyse activity dependence of critical percolation statistics in two different experimental model systems that realise distinct microscopic mechanisms of activity (Fig.~\ref{fig:exp}).
\rev{Although the difference between high and low activity is not visually detectable in the velocity and vorticity fields, we will show that they have fundamentally different symmetries.}
The first experimental system is the canonical two-dimensional active nematic of microtubule–kinesin motor protein mixtures~\cite{sanchez_spontaneous_2012, doostmohammadi2018active}, where activity can be tuned by varying the ATP concentration (Fig.~\ref{fig:exp}A, see details in \textbf{Methods}, section~\ref{sec:mtkin}, and full data in \textbf{Fig.\ref{fig:ED3}}).
At high ATP levels \rev{($[ATP] > \SI{18}{\micro\molar}$, calculations in Methods, section \ref{sec:mt_estim})}, the extracted vorticity fields exhibit nodal lines (zero-vorticity contours, see \textbf{Fig.\ref{fig:ED1}}) consistent with $SLE_6$: the left passage probability follows the analytical prediction (Fig.~\ref{fig:exp}C), and the Loewner driving function behaves as a \rev{Brownian (Wiener) process, therefore having a Gaussian distribution, with} diffusivity $\kappa \simeq 6$ (Fig.~\ref{fig:exp}D). 
At reduced ATP concentration \rev{($[ATP]< \SI{8}{\micro\molar}$)} both signatures are suppressed, indicating a breakdown of critical percolation statistics and the associated conformal symmetry.

The second system consists of dense suspensions of swimming bacteria, \rev{in which oxygen availability and the level of self-propulsion control the measured velocity} (Fig.~\ref{fig:exp}B, see details in  \textbf{Methods}, section~\ref{sec:bact}). 
Under high oxygen conditions, the flow fields again display critical percolation statistics with $SLE_6$ signatures, whereas depletion of oxygen leads to deviations from this behaviour (Fig.~\ref{fig:exp}C,D; \rrev{ED4}). 
\rev{To estimate the critical oxygen levels we compare the observed velocities to measurements with controlled oxygen concentration in~\cite{sokolov_physical_2012} and find that the transition point lies between $\SI{0.07}{\milli\molar}$ and $\SI{0.125}{\milli\molar}$ (see details in \textbf{Methods}, section \ref{sec:bact_estim}}).
Thus, in both experimental systems, critical percolation statistics emerges only at sufficiently high activity and disappears when activity is reduced.

\begin{figure}[t]
\centering
\includegraphics[width=\linewidth]{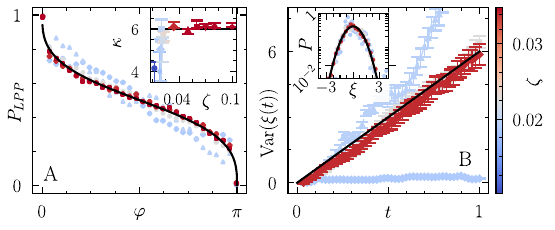}
\caption{\textbf{Activity–controlled emergence of critical percolation statistics in active nematic model.}
\textbf{(A)} Left–passage probability $P_{LPP}$ of nodal lines for different activity strengths~$\zeta$, indicated by colour, compared with the analytical Schramm–Loewner Evolution prediction for $SLE_{6}$ (black curve). 
(Inset) extracted diffusivity $\kappa$ from the left–passage statistics, showing convergence to $\kappa\simeq 6$ only above a critical activity $\zeta_{c}$\rrev{, with errors as defined in equation \ref{eq:lpperror}.}
\textbf{(B)} Variance of the Loewner driving function, ${\rm Var}(\xi(t))$. 
At high activity, the variance grows linearly in time with slope $\kappa\simeq 6$, consistent with Brownian statistics and $SLE_{6}$. 
At low activity, the variance deviates from linearity, indicating loss of conformal invariance. 
\rrev{Data as mean $\pm$ S.E.M.}
Inset: normalised distribution of $\xi(t)$ demonstrating Gaussian statistics only in the high–activity regime.
\rrev{For (A) and (B), $N=51, \zeta<0.019; N=401, \zeta\in[0.019, 0.022]; N=181, \zeta>0.022$, in order to resolve the transition region particularly clearly.}}
\label{fig:AN}
\end{figure}

\textit{Emergence and breaking of critical percolation statistics in active nematics.}—
To uncover the origin of this activity dependence, we turn to large-scale simulations of the hydrodynamic active nematic model~\cite{marchetti2013hydrodynamics,doostmohammadi2018active}, where a nematic order parameter $\mathbf{Q}$ couples to the flow through the active stress $\bm{\sigma}^{\mathrm{a}} = -\zeta \mathbf{Q}$, with activity parameter $\zeta$ (see details in \textbf{Methods}, section~\ref{sec:AN}). 
At high activity, the flow organises into a continuously evolving network of vortices and jets. 
In this regime, the nodal lines display the statistical signatures of critical percolation: the left passage probability follows the analytical prediction for Schramm–Loewner Evolution with diffusivity $\kappa = 6$ (Fig.~\ref{fig:AN}A), and the variance of the extracted Loewner driving function grows linearly in time with the same diffusivity $\kappa$ (Fig.~\ref{fig:AN}B). 
These independent measures confirm that the high-activity state of the active nematic is consistent with $SLE_6$, placing it in the universality class of two-dimensional critical percolation \rev{(see also \textbf{Methods}, section \ref{sec:cpvsan} and \textbf{Fig.\ref{fig:ED5}}}).

Reducing the activity below a threshold systematically breaks these signatures. 
The left passage probability deviates from the $SLE_{6}$ form and the driving function loses its Brownian character, demonstrating that the flow no longer exhibits critical percolation statistics. This loss is not merely a change in numerical exponents or curve roughness but signals a deeper transformation in the underlying geometric structure of the flow. Within the SLE framework, critical percolation statistics are inseparable from a specific symmetry principle, namely conformal invariance.

\rev{It is important to recall that conformal invariance is a highly restrictive symmetry, including translational, rotational, scale invariance, and invariance to angle preserving transformations (invariance to the special conformal transformation), which leads to tight constraints.}

In two dimensions, critical percolation provides a distinguished example of a conformally invariant state. Its interfaces are described by $SLE_{6}$, reflecting invariance under arbitrary conformal mappings. The observation that the nodal lines of the vorticity field obey $SLE_{6}$ therefore identifies the high activity phase with the universality class of critical percolation and establishes the presence of conformal symmetry in the flow geometry.

Conversely, when activity is reduced and the contour statistics deviate from $SLE_{6}$, this geometric universality is lost. Within the SLE framework, the breakdown of critical percolation statistics corresponds directly to the breaking of conformal invariance. 
The activity controlled crossover observed here thus represents a genuine symmetry breaking transition, with the symmetry gained or lost being conformal, imposing exceptionally strong constraints on the flow geometry across scales.

\begin{figure}[t]
\centering
\includegraphics[width=.99\linewidth]{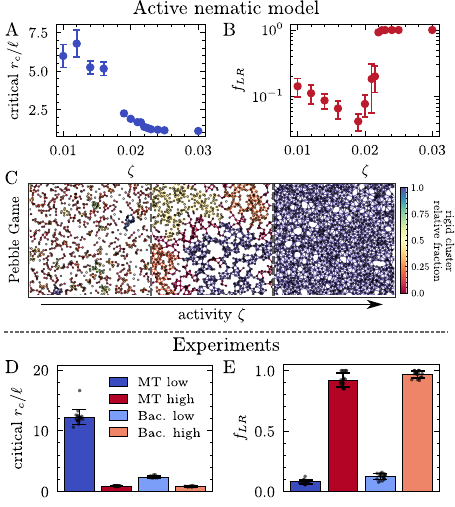}
\caption{\textbf{Activity-dependent geometric and mechanical percolation in simulations and experiments.}
\textbf{(A)} Critical connection length $r_c/\ell$ for vortex \rev{connectivity} percolation in active nematic simulations, defined by $r_c = r(P_{\infty}=0.5)$, where $\ell$ is the first peak of the pair correlation function $g(r)$.
\textbf{(B)} Fraction of vortices in the largest rigid cluster $f_{LR}$, showing a rigidity percolation transition at $\zeta_c$.
\textbf{(C)} Representative rigidity analysis from the Pebble Game algorithm, illustrating the emergence of a system-spanning rigid cluster only at high activity.
\textbf{(D-E)} Corresponding results for microtubule–kinesin active nematics and bacterial suspensions, both exhibiting reduced critical $r/\ell$ and increased $f_{LR}$ at high activity, consistent with simulations. Error bars indicate mean ± S.D. \rrev{For (A) and (B), $N=25$. For (D) and (E), the number of frames analysed are $N=18$ for all groups.}}
\label{fig:connectivity}
\end{figure}

\textit{Role of non-equilibrium fluctuations.}—
To test whether the establishment of $SLE_6$, and therefore critical percolation, requires a specific form of active stress or arises generically as a consequence of activity, we next replaced the active stress with stochastic forcing in a fluctuating nematohydrodynamic model~\cite{bonn2022fluctuation} (see details in \textbf{Methods}, section~\ref{sec:FN}). 
Here, the nematic tensor evolves under passive elastic stresses and viscous coupling, while random velocity and orientation fluctuations inject energy at small scales and explicitly violate detailed balance. 

In the presence of these non-equilibrium fluctuations, the system exhibits scale invariance of the type expected in $SLE_6$, with a fractal dimension of $d_f=7/4$ fulfilling critical percolation statistics (\textbf{Fig.~\ref{fig:ED6}A}, details in \textbf{Methods}, section~\ref{sec:fluctnem}).
Furthermore, measuring conformal invariance using the winding angle approach, we find that the active fluctuations lead to a winding angle scaling consistent with $SLE_6$ (\textbf{Fig.~\ref{fig:ED6}B}).
Finally, we can confirm $SLE_6$ behaviour by direct driving function measurement.

In contrast, when the same fluctuations are constrained to satisfy the fluctuation–dissipation relation, already scale invariance, a constituent symmetry of conformal symmetry, is broken and the more restrictive $SLE_6$ symmetry is not achieved (\textbf{Fig.~\ref{fig:ED6}A, B}).
The emergence of conformal invariance is therefore a genuine non-equilibrium effect, arising purely from breaking detailed balance, without the need for a specific form of active stress. In the following, we show that this symmetry lost/gained is accompanied by a simultaneous reorganisation of the vortex population at the mechanical level, revealing a deep connection between geometric criticality and mechanical constraints in active flows.\\

\textit{Rigidity percolation of vortex network.}—
To move beyond the statistics of nodal lines, we next examine how the underlying vortex structure reorganises as the system crosses the activity-driven transition associated with critical percolation statistics and the emergence of conformal symmetry.
Whereas the previous analysis focused on the geometry of large scale nodal lines contours which are the boundaries of regions of same sign vorticity, here we identify the discrete centres of these vortices and analyse how the \rev{instantaneous vortex configurations} connect and constrain one another.

Vortex cores are detected directly from the winding of the velocity field~\cite{hoffmann2021robustness} (\textbf{Fig.~\ref{fig:ED7}A}, see details in \textbf{Methods}, section~\ref{sec:vort_detect}). 
From their positions, we construct geometric graphs in which any pair of vortices separated by less than a distance $r$ are considered connected~\cite{pike1974percolation,penrose2003random}. 
\rev{We progressively increase the linking distance $r/\ell$, where $\ell = r(g_{\max})$ is defined by the position of the maximum of the vortex pair correlation function $g$ (\textbf{Fig.~\ref{fig:ED7}B}), and measure the probability $p_\infty$ of forming a spanning cluster (\textbf{Fig.~\ref{fig:ED8}}). 
At low activity, the percolation threshold, defined by $p_\infty = 0.5$, is crossed only at large values of $r$, with $r/\ell \gg 1$, indicating the absence of an intrinsic length scale. 
In contrast, above the transition, $\zeta > \zeta_c$, the curve $P_\infty(r/\ell)$ becomes sharp, and percolation occurs at $r/\ell \approx 1$, showing that $\ell$ emerges as the natural length scale controlling connectivity in the system.}
This analysis reveals that a system-spanning cluster forms only above the same critical activity $\zeta_c$ identified from the conformal-symmetry diagnostics: at low activity, the vortices remain isolated or form small groups, whereas at high activity they merge into a single connected cluster that spans the entire system. 

Geometric connectivity alone, however, does not capture the mechanical integrity of the vortices' spatial organisation. 
We therefore constructed networks of vortices and performed rigidity percolation analysis using the Pebble Game algorithm~\cite{jacobs1995generic, petridou2021rigidity}.
At each activity, we fix the connection threshold \rev{$r = 2\ell$} and generate alpha shapes~\cite{Edelsbrunner1994Three-dimensionalShapes, edelsbrunner1993union}---embedded graphs in which pairs of vortices are connected if they are within a distance $r$, provided the connecting edge does not intersect others (see details in \textbf{Methods}, section~\ref{sec:mech_perc}).
We then identify rigid sub-clusters within the network and calculate the relative size of the largest rigid sub-cluster $f_{LR}$~\cite{jacobs1995generic,jacobs1997algorithm} (Fig.~\ref{fig:connectivity}B,C).
The results show that, precisely at $\zeta_c$, a giant rigid, mechanically constrained cluster spanning the system ($f_{LR}\rightarrow1$) emerges (Fig.~\ref{fig:connectivity}B): the system transitions from a loosely connected collection of vortices to a rigid, self-supporting flow network. We also performed the same geometric and mechanical analyses on microtubule–kinesin motor protein mixtures and dense suspensions of swimming bacteria (Fig.~\ref{fig:connectivity}D,E).
In both cases, we observe the same trend: high-activity systems exhibit larger geometric clusters and mechanically rigid clusters compared to their low-activity counterparts.

These analyses provide a direct geometric and mechanical picture of the activity-driven phase transition. 
They demonstrate that the onset of critical percolation statistics of nodal lines coincides with a collective reorganisation of the vortex population, where activity concentrates rotational energy into a percolating, rigid structure. 
\rev{The interfaces of this structure correspond to the conformally invariant nodal lines, linking the flow geometry and its mechanical backbone through a common critical threshold.}\\

\textit{Topology of the vorticity landscape.}—
The geometric and mechanical analyses above show how activity reorganises the vortex population.
We now move beyond these discrete structures to examine the topology of the vorticity field itself, asking whether the field-level organisation undergoes a corresponding transition.
We use persistent homology~\cite{Verri1993, Robins1999, Edelsbrunner2002}, a method that quantifies how topological features, \textit{i.e.} connected components and loops in the vorticity landscape, appear and disappear as we probe the field's topology at different length scales (See \textbf{Methods}, section~\ref{sec:PH} and schematic~\ref{fig:ED9}A, B).
Unlike contour-based or vortex-based measures, this approach captures the global structure of the entire vorticity field and is sensitive to multi-scale organisation. \rev{In this sense, it provides an intermediate description between the geometry of nodal lines and the discrete network of vortex centres, by tracking how structures evolve continuously across spatial scales.}

\begin{figure}[ht]
    \centering
    \includegraphics[width=\linewidth]{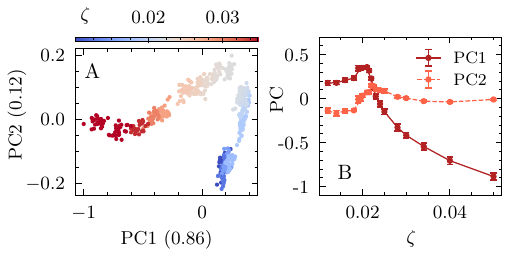}
    \caption{\textbf{Topological restructuring of the vorticity field at the transition activity threshold.}
    \textbf{(A)} Principal component analysis of persistence images, with fraction of explained variance in parentheses, reveals a clear separation between conformally invariant and symmetry-broken states, with the first component capturing the dominant variation across activities.
    \textbf{(B)} Projection onto the first two components highlights the transition as a distinct shift in the topology of the vorticity landscape.
    \rrev{Data as mean $\pm$ standard deviation. $N=51, \zeta<0.019; N=401, \zeta\in[0.019, 0.022]; N=181, \zeta>0.022$, in order to resolve the transition region particularly clearly.}}
    \label{fig:PH}
\end{figure}

For each activity, we compute the first persistence diagrams of the signed Euclidean distance transformed (SEDT) vorticity field~\cite{Obayashi2022}, which capture the birth and death of features as we flood the SEDT landscape, and convert them to persistence images~\cite{Adams2015} (see Fig. \textbf{\ref{fig:ED9}}). \rev{Intuitively, this procedure can be viewed as gradually “flooding” the vorticity landscape: \rrev{ when the water level is very low, the whole landscape is connected, but as the water level rises the landscape breaks up into islands and finally only mountain tops are visible until everything is flooded. 
For each topological feature, where an island is the $0$th homology and a lake (a hole in an island) is the $1$st homology, we record the time of birth and the time of death.
Therefore,} smooth, weakly structured flows produce features with similar lifetimes, while more heterogeneous and intermittent flows generate a broader distribution of lifetimes. 
\rrev{A simple lifetime histogram already clearly shows the transition from a population with a clear mean lifetime to a wider distribution of lifetimes~\ref{fig:ED10}.}
Since these persistence images are high-dimensional objects (Fig. \textbf{\ref{fig:ED9}}B), we use principal component analysis to extract the dominant modes of variation and obtain a low-dimensional representation that can be directly compared across activity levels.}
Principal component analysis of these images reveals a clear separation between low-activity and high-activity regimes, with a transition at the same critical activity $\zeta_c$ identified from the SLE diagnostics and vortex-network analyses (Fig.~\ref{fig:PH}).
At this threshold, the vorticity landscape reaches its most distinct topological configuration.

Persistent homology therefore provides a third, independent, field-level diagnostic of the phase transition.
The consistent identification of the same activity threshold---from nodal-line geometry, vortex-network connectivity, and vorticity field topology---demonstrates that the transition is a robust, system-wide reorganisation of the active flow.

\textit{Discussion.—}
Our results show that active fluids undergo an activity-driven transition in which critical percolation statistics emerge in the geometry of the flow. Once activity exceeds a well defined threshold, the nodal lines of the vorticity field form a system spanning backbone with the statistics of critical percolation.
By combining complementary analyses of nodal lines, vortex connectivity, and field level topology, we demonstrate that the same activity threshold marks a collective reorganisation of symmetry, geometry, and mechanics. The convergence of these independent signatures, from nodal lines to vortex centres and the continuous vorticity field, provides a stringent and quantitative criterion for identifying fully developed active turbulence. We establish this behaviour in two distinct experimental systems and reproduce it in simulations of active nematics and fluctuating nematics, showing that it is an intrinsically non-equilibrium phenomenon.

The emergence of critical percolation statistics has a direct symmetry interpretation. In two dimensions, critical percolation is a distinguished conformally invariant state, whose interfaces are described by Schramm Loewner Evolution with diffusivity $\kappa = 6$ ($SLE_6$).
The observation of $SLE_6$ statistics therefore implies the emergence of full conformal symmetry in the flow geometry. 
Conversely, when activity is reduced and these statistics break down, conformal symmetry is lost. The transition identified here is thus a symmetry breaking transition, but of an unusual kind, where the symmetry gained or lost is conformal. This places exceptionally strong constraints on the geometry of fluctuations across scales and elevates fully developed active turbulence to a non-equilibrium critical state, rather than a merely disordered regime.

Beyond geometry, we find that the same activity threshold controls the mechanical organisation of the flow. Vortex centres undergo rigidity percolation at the transition, forming a mechanically constrained network that spans the system.
The coincidence of geometric critical percolation and mechanical rigidity percolation reveals that increasing activity does not simply enhance disorder, but instead imposes collective mechanical constraints for large scale flow structures. 
This identifies a concrete mechanical backbone underlying the critical geometry of the vorticity field.

\rev{This perspective also helps connect our results to more directly observable and functional aspects of active systems. In many biological and active materials, changes in flow organisation are linked to changes in transport, mixing, and force transmission~\cite{sanchez_spontaneous_2012,serra2023defect,ma2025cellular,zhao2024asymmetric,monfared2024short}. In this context, the transition we identify can be viewed as the point at which the flow becomes coordinated across the entire system, both geometrically and mechanically. The emergence of a system spanning structure therefore provides a simple and physically intuitive picture for how increasing activity can reorganise the flow into a collective state.}

The connection to $SLE_6$ places fully developed active turbulence alongside a growing class of non-equilibrium systems where conformal invariance has recently been identified, including driven amorphous solids, water wave turbulence, and liquid crystal interfaces~\cite{javerzat_schramm-loewner_2024,noseda2024conformal,almeida_critical_2025}. In parallel, recent theoretical efforts have begun to bridge conformal symmetry, topology, and non-equilibrium field theory, through effective action approaches for active matter that incorporate modified dynamical KMS symmetry~\cite{landry2023active}, geometric formulations of coarse grained active flows~\cite{armas2025hydrodynamics}, and topological descriptions of nodal structures in two dimensional turbulence~\cite{eling2025topology}. Together with the experimental and numerical results presented here, these developments suggest that a unified theory of non-equilibrium conformal invariance may now be within reach.

Several open questions follow naturally from our findings. Most striking is the coincidence between critical percolation of nodal lines and rigidity percolation of vortex centres, despite their distinct universality classes in equilibrium settings. This suggests a previously unexplored link between geometric criticality and mechanical constraints in driven fluids. \rev{Moreover, while in our modelling here we only analysed active nematohydrodynamics as a continuum basis for active turbulence, other continuum models that explicitly account for polar effects of self-propulsion~\cite{wensink_meso-scale_2012,linkmann_condensate_2020,worlitzer2021motility,backofen2024nonequilibrium,amiri2022unifying,de2025self} can be analysed within the same framework. Future work should assess similarities and differences of these different continuum models in terms of the reported transition here.} More broadly, our fluctuating nematohydrodynamic simulations show that explicit active stresses are not required, as violation of detailed balance alone can generate conformally invariant states. This points to conformal symmetry as a generic organising principle of non-equilibrium dynamics, and not a property that is tied to any specific microscopic mechanism.

\section*{Acknowledgements}
We thank Mehran Kardar for helpful discussions.
The Tycho supercomputer hosted at the SCIENCE HPC center at the University of Copenhagen was used for supporting this work. 
\section*{Funding statement}
M.C.P. thanks Villum Fonden (Grant No.~69081) for their support.
A.D. acknowledges funding from the Novo Nordisk Foundation (grant No.~NNF18SA0035142 and NERD grant No.~NNF21OC0068687), Villum Fonden (Grant No.~29476), and the European Union (ERC, PhysCoMeT, 101041418).
NA acknowledges financial support from the Portuguese Foundation for Science and Technology (FCT) under Contracts no. UID/00618/2025, UID/PRR2/00618/2025, and UID/PRR/00618/2025.
The authors are indebted to the Brandeis University MRSEC Biosynthesis facility for providing the tubulin. We thank M. Pons and A. Fernández (Universitat de Barcelona) for their assistance in the expression of motor proteins. O. B., J.I.-M., and F.S. acknowledge funding from MICINN/AEI/10.13039/501100011033 (Grant No. PID2022-137713NB-C21). O. B. acknowledges a Joan Oró FI fellowship (2023 FI-3 00065) from Generalitat de Catalunya and the European Social Plus Fund. Brandeis University MRSEC Biosynthesis facility is supported by NSF MRSEC 2011846. G.J. acknowledges financial support from the National Natural Science Foundation of China (W2421001, 12174306), and Natural Science Basic Research Program of Shaanxi (2023-JC-JQ-02).
\section*{Author Contributions Statement}
A.D. and N.A.M.A. designed the study.
O.B. designed and performed the microtubule kinesin experiments with variable activity and the PIV analysis.
O.B., J.I.M., and F. S. developed the ATP concentration calculation.
W.F. and G.J. designed and performed the bacterial experiments.
L.B. and T.M. performed SLE analysis on the experimental data.
L.B. performed SLE analysis on the simulation data.
T.M. performed the vortex centre analysis.
M.C.P. performed the PH analysis.
L.B. wrote the first draft.
L.B., T.M., and A.D. wrote the paper.
G.J., J.I.M., F.S., N.A.M.A., and A.D. designed the experiments and models, as well as the interpreting results.
A.D. led the collaborative effort.

\section*{Competing Interests Statement}
The authors declare no competing interests
\section*{References}

\newpage
\section*{Methods}

\counterwithin*{figure}{part}

\stepcounter{part}

\renewcommand{\thefigure}{ED\arabic{figure}}


\renewcommand\thetable{M\arabic{table}}    
\setcounter{table}{0}   

\section{Schramm Loewner evolution details} \label{sec:SLESI}

Schramm Loewner Evolution (SLE) is a one-parameter family of conformally invariant random curves~\cite{cardy_sle_2005, gruzberg_loewner_2004}.
The single parameter $\kappa$ controls the fractal dimension of the curve, and specific values of $\kappa$ correspond to the scaling limits of several well-known models: the loop-erased random walk with $\kappa=2$~\cite{lawler2011conformal}, the Ising model with $\kappa=3$~\cite{chelkak_convergence_2014}, and critical percolation with $\kappa=6$~\cite{smirnov2001critical}.
This makes SLE a powerful framework for identifying universality classes.

The Loewner equation describes the evolution of a conformal map $g_t$ in Loewner time $t$:
\begin{equation}
    \frac{d}{dt} g_t (z)= \frac{2}{g_t(z) - \xi_t},\quad g_0(z) = z,
\end{equation}
where $\xi(t)$ is the driving function with Gaussian statistics $\langle\xi_t\rangle = 0$, $\langle \xi_t\xi_{t'}\rangle = \kappa\delta(t-t')$.
The curve $\gamma$, known as the SLE trace, is traced out by those points satisfying $g_t(z_c(t))=\xi_t$, where $z_c$ is the preimage of the real line.

By analogy with Brownian motion, $\kappa$ acts as a diffusivity that determines the strength of the driving in the Loewner equation.
For $\kappa=0$, the curve $\gamma$ is a line with fractal dimension $d_f=1$, while increasing $\kappa$ makes $\gamma$ progressively more winding or craggy until it fills the plane, reaching $d_f=2$, according to the relation $d_f = \min(2, 1+\kappa/8)$~\cite{beffara_dimension_2008}.

To determine $\kappa$ and thus identify the universality class of our curves, we use two independent methods: the left passage probability and the driving function.

\begin{figure}[ht]
    \centering
    \includegraphics[width=0.8\linewidth]{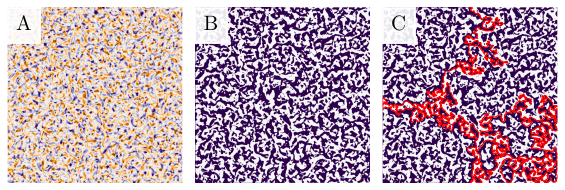}
    \caption{\textbf{Extraction of the nodal line  (zero-vorticity isoline)}
    (A) Vorticity field of active nematics in the fully developed turbulent regime. 
    (B) Binarised field showing regions of positive and negative vorticity. The boundary of left half of the field is set to $1$ and the boundary right half to $0$, ensuring that the isoline starts and ends at midpoints on the bottom and top boundaries, respectively, without exiting the field.
    (C) The resulting zero-vorticity isoline follows the interface between positive and negative regions, starting at $x=L/2$, $y=0$ and terminating at $x=L/2$, $y=L$.}
    \label{fig:ED1}
\end{figure}

\paragraph{Extracting candidate traces}
To extract the nodal line (zero-vorticity isoline), we follow the standard procedure as in~\cite{andersen_evidence_2025}.
Starting from the vorticity field (Fig.~\ref{fig:ED1}A), we binarise it as {\texttt{b = $\omega$ > 0}}, rotate by $\pi/2$ with a probability of $p=0.5$, and set the origin at the centre of the lower boundary.
At the domain edges, we assign value $1$ for $x < 0$ and $0$ otherwise (Fig.~\ref{fig:ED1}B), ensuring that the trace starts at $(x=0, y=0)$ and ends at $(x=0, y=L)$.
An explorer is then initiated at the origin and follows the boundary between $0$ and $1$, keeping $1$ to its left (Fig.~\ref{fig:ED1}C).

\begin{figure}
    \centering
    \includegraphics[width=0.5\linewidth]{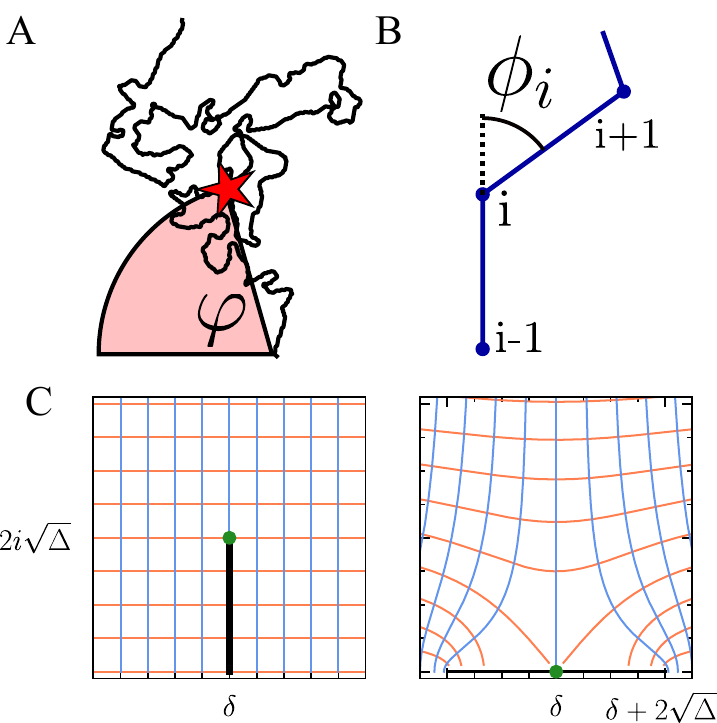}
    \caption{\rev{\textbf{Methods used to measure conformal invariance}
    \textbf{(A)} The left passage probability $P_{LPP} = P(\kappa, \varphi)$ is measured by measuring the probability of a point with polar angle $\varphi$ to lie left of the candidate trace.
    \textbf{(B)} Winding angle measurement consists of measuring the angle between two consecutive line segments of the discretised curve.
    \textbf{(C)} Illustration of the vertical slit map $g_t(z)$.
    On the left is the complex plane with the slit from $\delta$ to $2i\sqrt{\Delta} + \delta$ drawn in black. 
    On the right is the complex plane when $g_t$ has been applied: the slit has been mapped onto the real line from $\delta$ to $2\sqrt{\Delta} + \delta$ and the tip of the slit at $z= \delta + 2i\sqrt{\Delta}$ is mapped straight down to $g(z) = \delta$ (green dot).
    }
    }
    \label{fig:ED2}
\end{figure}

\paragraph{Left passage probability}
The left passage probability (LPP) is defined as the probability that a point in the plane lies to the \rev{left of the SLE trace $\gamma$, that is, that the curve passes right of a given point (\textbf{Fig.\ref{fig:ED2}A}).
In other words, left passage means that there is a path from a point $x$ on the real line with $Re(x)<0$ to our test point which does not cross the trace.}
It depends only on the polar angle $\varphi$ from the curve’s origin, and the analytical form is known~\cite{schramm_percolation_2001} (see Eq.~\eqref{eq:lpp} in the main text).
We compare the measured LPP for a set of points with cardinality $S$ to the analytical prediction and extract $\kappa$ through fitting.
The mean square deviation $Q(\kappa)$ is defined as
\begin{align}
    Q(\kappa) = \frac{1}{S}\sum_{z\in S}\frac{(p(\varphi) - P_\kappa(\varphi))^2}{p(\varphi)(1-p(\varphi))},
\end{align}
where $p(\varphi)$ is the measured probability and $P_\kappa(\varphi)$ is the theoretical distribution~\cite{norrenbrock_paths_2013, pose_shortest_2014}.
The reported diffusivity $\kappa^*$ corresponds to the minimum of $Q(\kappa)$.
The uncertainty range of $\kappa$ is then defined by the set of $\kappa \in K = [0, 8]$ values satisfying $\mathcal{K} = \{\kappa \in K | Q(\kappa^*) + \Delta Q(\kappa^*) \geq Q(\kappa) - \Delta Q(\kappa)\}$ where $\Delta Q$ is the measured uncertainty in the mean square deviation. 
The error bar is constructed as:
\begin{align}
   \min \mathcal{K}, \kappa^*, \max \mathcal{K} \label{eq:lpperror}
\end{align}

\paragraph{Driving function}
A more direct method to determine $\kappa$ is to measure the driving function $\xi_t$ itself.
We discretise Loewner time into steps $t_i$ with intervals $\Delta_i = t_i - t_{i-1}$ and approximate $\xi_t$ as constant within each interval, $\xi_{t_i} = \delta_i$.
We then apply the vertical slit map $g_{t_i} = \sqrt{(z-\delta_i)^2 + 4\Delta_i} + \delta_i$ \rev{\textbf{(Fig.\ref{fig:ED2})}}, which maps the vertical slit extending from $\delta$ to $\delta + 2i\sqrt{\Delta}$ onto the real axis, effectively “unzipping’’ the trace $\gamma$~\cite{kennedy_numerical_2009}.
At each time step, we record the value of $\xi$, and by averaging over many realizations of $\gamma$, we obtain the trajectory of $\xi_t$.
The diffusivity $\kappa$ is then determined from the variance $\langle \xi^2(t) \rangle = \kappa t$ and verified by checking that $\xi_t$ is Gaussian distributed.

\section{Experimental details}
\subsection{Microtubule–Kinesin system} \label{sec:mtkin}

\paragraph{Protein preparation} 
Stabilised microtubules (MTs) were polymerised from heterodimeric ($\alpha, \beta$)-tubulin purified from bovine brain (Biomaterials Facility, Brandeis University MRSEC, Waltham, MA) using a non-hydrolysable GTP analogue, guanosine-5-[($\alpha, \beta$)-methyleno]triphosphate (GMPCPP) (Jena Biosciences, NU-405), following a previously published protocol~\cite{tayar_assembling_2022}.  

A construct encoding the \textit{Drosophila melanogaster} heavy-chain kinesin-1 fragment (Addgene ID: 15960) was expressed in \textit{E. coli} Rosetta (DE3), and recombinant kinesin protein was purified according to established protocols~\cite{tayar_assembling_2022}.

\paragraph{Assembly of the active gel} 
Kinesin motor dimers were prepared by mixing biotinylated kinesin motor proteins with tetrameric streptavidin (Invitrogen, 434301) in a 2:1 molar ratio in the presence of 0.22 mM DL-dithiothreitol (DTT) (Sigma-Aldrich, 43815). The mixture was incubated on ice for 30 minutes.  

The dimerised motor complexes were then combined with a feeding solution containing ATP (Sigma, A2383), an ATP-regenerating system (phosphoenolpyruvate (PEP) (Sigma, P7127) and pyruvate kinase/lactate dehydrogenase (PK/LDH) (Sigma, P0294)), the nonadsorbing polymer poly(ethylene glycol) (PEG, 20 kDa) (Sigma, 95172) that promotes filament bundling through depletion, an oxygen scavenging and antioxidant system (catalase (Sigma-Aldrich, C40), glucose oxidase (Sigma-Aldrich, G2133), D-(+)-glucose (Sigma-Aldrich, G7021), Trolox (Sigma-Aldrich, 238813), and DTT), and the PEG-based triblock copolymer surfactant Pluronic F-127 (Sigma, P-2443).  

Microtubules were added to the mixture immediately before the experiment. The final concentrations of all components are listed in Table~\ref{tab:gel_mixture}.

\begin{table}[ht]
    \centering
    \caption{Composition of the active gel mixture.}
    \label{tab:gel_mixture}
    \begin{tabular}{|l|l|}
        \hline
        \textbf{Compound} & \textbf{Final concentration} \\
        \hline
        PEG (20 kDa)*       & 1.54\% w/v \\ \hline
        PEP*                & 25.68 mM \\ \hline
        MgCl$_2$*           & 3.12 mM \\ \hline
        Trolox**            & 1.93 mM \\ \hline
        ATP*                & 1.37 mM \\ \hline
        Catalase**          & 0.04 mg/ml \\ \hline
        Glucose**           & 3.20 mg/ml \\ \hline
        Glucose oxidase**   & 0.21 mg/ml \\ \hline
        PK/LDH              & 25.01 / 24.91 U/ml \\ \hline
        Pluronic F-127*     & 0.41\% w/v \\ \hline
        DTT*                & 5.21 mM \\ \hline
        Streptavidin*       & 0.01 mg/ml \\ \hline
        Kinesin*            & 0.08 mg/ml \\ \hline
        Microtubules*       & 1.85 mg/ml \\ \hline
    \end{tabular}

    \vspace{0.5em}
    \raggedright
    \footnotesize
    *M2B buffer: 80 mM PIPES pH 6.8, 2 mM MgCl$_2$, 1 mM EGTA.\\
    **Phosphate buffer: 6.68 mM KH$_2$PO$_4$, 12.32 mM K$_2$HPO$_4$, pH 7.2.
\end{table}

\paragraph{Reactivation solution} 
Over time, the aqueous phase above the inactive active-nematic (AN) layer at the water–oil interface accumulated enzymatic byproducts. This liquid was exchanged with a fresh reactivation solution containing all components required for active gel preparation (see Table~\ref{tab:gel_mixture}), except microtubules. 
ATP was replaced with the NPE-caged ATP (adenosine 5'-triphosphate, P3-(1-(2-nitrophenyl)ethyl) ester and disodium salt) (Thermo Fisher Scientific, A1048) in order to allow the use of UV light to control the release of ATP available to the kinesin motors.

\paragraph{Active nematic cell} 
Experiments were performed in flow cells with a channel width of 1.5–2 mm, length of 15 mm and height of 150 $\mu$m. Each cell was assembled from superhydrophilic polyacrylamide-coated glass and superhydrophobic Aquapel®-coated glass, separated by 150 $\mu$m thick double-sided tape.  
Before coating, two holes of 1.5–2 mm diameter and 15 mm separation were drilled with a diamond wheel point (Dremel, 7134). These openings were later used for filling and exchanging the aqueous phase with the reactivation solution.  

The cell was initially filled by capillarity with fluorinated oil (HFE7500, Fluorochem 051243) containing 2\% fluorosurfactant copolymer (RanBiotechnologies, 008 Fluorosurfactant). The active material was then introduced by capillarity, displacing the oil except for a thin lubricating layer. To prevent evaporation, the cell was sealed with petroleum jelly.  

The AN formed at the water–oil interface and became inactive after about 24 hours due to ATP depletion and accumulation of byproducts. To reactivate the AN, the aqueous phase above the inactive layer was replaced with the reactivation solution. The petroleum jelly was temporarily removed to expose the channel openings. A drop of reactivation solution (two to three times the channel volume) was added to one hole, and a Kimwipes® wiper (Kimtech, 34120) was inserted in the opposite hole to induce capillary flow.  
After replacement, the cell was resealed with petroleum jelly and illuminated with UV light (ThorLabs, M365LP1: 365 nm, 1350 mW LED) to release ATP from NPE-caged ATP.  
Illumination was applied in pulses (2 Hz, 5\% duty cycle, 8.09 mW/cm\textsuperscript{2}) for 5 or 10 seconds. With this procedure, we first preform an AN layer at the water–oil interface. After losing activity, this layer preserves nematicity and maintains its integrity, at least over the time scales of the experiment. Following the exchange procedure with the reactivation solution, we released ATP from NPE-caged ATP in a controlled manner. This allowed us to study AN flows at extremely low activity levels, followed by a gradual increase in activity until fully developed active turbulence was achieved.

\paragraph{Imaging of the active nematic} 
Samples were imaged by fluorescence microscopy. Microtubule fluorescence was excited with a white LED source (Thorlabs MWWHLP2) and a Cy5 filter set (Edmund Optics) and recorded using a CCD camera (ExiBlue, QImaging). Images were captured and processed using the open-source software $\mu$Manager (ImageJ).

\paragraph{Image analysis} 
Raw experimental images were preprocessed in ImageJ and analyzed using PIVLab in MATLAB to extract and quantify velocity fields~\cite{thielicke_particle_2021}.

\rev{
\subsection{Microtubule-Kinesin activity estimation} \label{sec:mt_estim}
In this section we estimate the ATP concentration in the samples, allowing us to define the regions of low and high activity and to confine the conformal phase transition to activities in the interval $[ATP]_c \in (8, 18)\,\SI{}{\micro\molar}$.
}
\rev{
We assume that UV illumination is spatially uniform across the entire experimental cell and that NPE–caged ATP is homogeneously distributed within the sample.
Reflections of UV light from the output glass window back into the sample are neglected.
To evaluate the activity of the AN, we estimated the concentration of ATP released after photolysis of NPE-caged ATP as follows.
The fraction of absorbed light, $f_{abs} = 1-10^{-\mathcal{A}}$, was calculated from the absorbance of the reaction mixture, $\mathcal{A}$, using the Lambert–Beer law with the extinction coefficient for NPE-caged ATP, $\varepsilon_{\SI{355}{\nano\metre}} = \SI{430}{\litre\per\mol\per\centi\metre}$\cite{josts_photocage-initiated_2018}. 
Then, to estimate the effective photocleavage of NPE-caged ATP, we first calculated the number of photons reaching the sample per second based on power measurements of the UV light at the sample surface after transmission through the glass slide. 
To achieve this, the photodiode sensor (Thorlabs, S120VC) was positioned precisely at the sample location to maintain the geometrical fidelity of the illumination profile. 
The detector surface was covered with a glass slide, reproducing the optical interface present in experiments, and was exposed to identical incident power levels as those applied in the experiments.
The number of photons reaching the sample per second was calculated as $$N_{photons/s} = \frac{P\frac{A_{cell}}{A_{det}}}{E_{\SI{356}{\nano\metre}}},$$ where $P = \SI{8.09}{\milli\watt}$ is the UV light power reaching the detector,  $A_{det} = \SI{0.96}{\square\centi\metre}$ and $A_{cell} = \SI{0.2625}{\square\centi\metre}$ are the areas of the detector and the experimental cell, respectively, and $E_{\SI{356}{\nano\metre}} = \SI{5.446e-19}{\joule}$ is the energy of a UV photon.
}
\rev{
Using this value, the effective illumination time, and the fraction of absorbed light, we estimated the number of absorbed photons. After correcting for the quantum yield of NPE-caged ATP, $\phi_{360} = 0.6$ \cite{josts_photocage-initiated_2018, mccray_new_1980}, we estimated the number of ATP molecules released and thus their concentration in the volume of the experimental cell.
}
\rev{
The estimated ATP concentrations correlate with the \rrev{mean enstrophy $e$ obtained from PIV analysis: low-activity experiments occur below $[ATP] \approx\SI{8}{\micro\molar}$ and $e \approx \SI{2.8\pm0.1}{\per\square\second}$, while high-activity experiments occur above $[ATP] \approx \SI{18}{\micro\molar}$ and $e \approx \SI{4.1\pm0.4}{\per\square\second}$.
The measured transition ATP concentration does not coincide with the increase in shear rate in microtubule kinesin systems measured in \cite{lemma_multiscale_2021}.
}}
\subsection{Bacterial suspension system} \label{sec:bact}

Dense suspensions of flagellar-propelled bacteria were confined within quasi-two-dimensional wells, where collective motion spontaneously emerged and evolved dynamically as oxygen was depleted.  
\textit{Bacillus subtilis} (strain 168), a rod-shaped bacterium, was used as the self-propelled microswimmer.  

Cells were revived from frozen glycerol stocks stored at $\SI{-80}{\celsius}$ and inoculated into 10 mL of standard Luria–Bertani (LB) medium containing 1.0\% tryptone, 0.5\% yeast extract, and 1.0\% NaCl. Cultures were incubated overnight at 30 °C with shaking at 200 rpm.  
An aliquot of the overnight culture was then diluted into fresh LB to an initial optical density OD$_{600} \approx 0.05$ and grown for 6–7 hours to mid-exponential phase (OD$_{600} \approx 0.6$).  
Cells were harvested at 3000$\,g$ for 5 minutes, washed twice with motility buffer (10 mM potassium phosphate, 0.1 mM EDTA, 10 mM NaCl, pH 7.0), and resuspended in the same buffer for experiments at a final OD$_{600} \approx 72$.  
The typical bacterial body length was about $7~\mu$m and diameter about $1~\mu$m. Note that the bacterial volume fraction is only approximately estimated as $\phi \simeq 0.072$, since an accurate determination is non-trivial. Typically, an optical density of $\mathrm{OD}_{600}=1$ corresponds to approximately $3\times10^{8}$ bacteria per millilitre. Assuming a characteristic volume of $\sim 1~\mu\mathrm{m}^3$ per bacterium, this estimate yields a volume fraction on the order of $0.1\%$

Quasi-two-dimensional confinement was achieved using a thin layer of polydimethylsiloxane (PDMS). A PDMS sheet of size 1 cm $\times$ 1 cm was patterned with an array of circular wells (diameter 500~$\mu$m, depth 10~$\mu$m).  
The PDMS wells were plasma-treated to render them hydrophilic and able to hold the bacterial suspension.  
A 2~$\mu$L aliquot of concentrated bacterial suspension was deposited onto a treated cover glass and overlaid with the PDMS structure to form a sealed observation chamber.  
The assembled glass–PDMS construct was placed in a humidity-controlled environment with relative humidity above 90\%.  

Bacteria near the glass interface in the circular wells were imaged using a Nikon Ti2-E inverted microscope equipped with a 60× water-immersion objective (NA = 1.2) and a high-speed camera (Hamamatsu, ORCA-Flash4.0 V3).  
As external oxygen was cut off, its concentration gradually decreased due to bacterial consumption, leading to a slow decay of collective activity.  
Videos were recorded every five minutes, capturing the progressive loss of activity. Each recording was taken at 50 frames per second, and every third frame was used for PIV analysis to ensure reliable velocity field estimation.  

\begin{figure}
    \centering
    \includegraphics[width=0.5\linewidth]{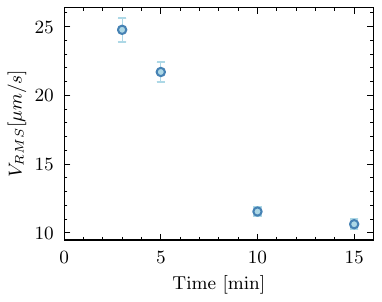}
    \caption{
    \rev{
    \textbf{Temporal evolution of bacterial velocity.} Bacterial $V_\text{RMS}$ decreases with time in the closed system.}
    \rrev{$N_{\SI{3}{\minute}} = 105, N_{\SI{5}{\minute}} = 250, N_{10, 15\, \SI{}{\minute}} = 50$. Data are presented as mean $\pm$ S.D.}
    }
    \label{fig:ED4}
\end{figure}

\rev{
\subsection{Bacterial activity estimation} \label{sec:bact_estim}
Having found a transition between high and low activity in bacterial systems, we wanted to understand the transition point more quantitatively.
Therefore, we compare our experiments to previous experiments on bacteria where the oxygen levels were varied.
Starting from the observation that the root mean squared velocity $V_\text{RMS}$ measured by PIV decreases over time, and that at some point during this process $SLE_6$ breaks, we wanted to understand how the activity changes in that time (see \textbf{Fig.\ref{fig:ED4}}). 
Experiments on bacterial collective velocity, with controlled oxygen conditions, show that the relationship between oxygen concentration and collective velocity is linear~\cite{sokolov_physical_2012}.
We therefore compare our system to a system that has controlled oxygen levels.
We note that our system starts at $\SI{25}{\celsius}$ and standard atmospheric pressure, leading to a concentration of oxygen in water equilibrated with air of $\SI{0.25}{\milli\molar}$.
So, at $t=0$ and therefore oxygen concentration of $\SI{0.25}{\milli\molar}$ we find $V_\text{RMS}(t=\SI{3}{\minute}) \approx \SI{25}{\micro\metre\per\second}$ (\textbf{Fig.\ref{fig:ED4}}) which aligns well with previous results~\cite{sokolov_physical_2012}.
We then find that the activity drops with time, while time points $\SI{3}{\minute}$ and $\SI{5}{\minute}$ show $SLE_6$, time points $\SI{10}{\minute}$ and $\SI{15}{\minute}$ show broken $SLE_6$.
This means that the transition occurs between $\SI{12}{\micro\metre\per\second}$ and $\SI{21.7}{\micro\metre\per\second}$, which, comparing with~\cite{sokolov_physical_2012}, corresponds to a transition at oxygen concentrations between $\SI{0.07}{\milli\molar}$ and $\SI{0.125}{\milli\molar}$.
}


\section{Model descriptions and computational details} \label{sec:models}

\subsection{Active nematics} \label{sec:AN}
We use a standard active nematics model as described in~\cite{gennes_physics_1993, marchetti2013hydrodynamics, thampi_active_2016, doostmohammadi2018active}.
We begin by defining a Landau de Gennes free energy with bulk constant $C$ and a Frank elastic term with elastic constant $K$ in terms of the order parameter $\mathbf{Q}$:
\begin{equation}
    \mathcal{F} = \int d\vec r\, C(1-\tr(\mathbf{Q}^2))\tr(\mathbf{Q}^2) + \frac{K}{2}(\nabla\cdot\mathbf{Q})^2.
\end{equation}

From this free energy, we derive the molecular field $\mathbf{H} = \left(\frac{\delta\mathcal{F}}{\delta\mathbf{Q}}\right)^{ST}$, which is the symmetric and traceless (ST) part of the functional derivative of the free energy with respect to the order parameter $\mathbf{Q}$.

The dynamics of $\mathbf{Q}$ is governed by its relaxation towards the free energy minimum and by its coupling to the flow field $\vec u$:
\begin{equation}
    \frac{D \mathbf{Q}}{Dt} - \mathbf{S} = -\frac{1}{\gamma}\mathbf{H},
\end{equation}
where $D/Dt = \partial_t + \vec{u}\cdot\nabla$ is the material derivative.
The corotation term $\mathbf{S} = (\lambda\mathbf{E}+\mathbf{\Omega})\cdot(\mathbf{Q}+\mathbf{I}/2) + (\mathbf{Q}+\mathbf{I}/2)\cdot(\lambda\mathbf{E}-\mathbf{\Omega}) - 2\lambda(\mathbf{Q}+\mathbf{I}/2)(\mathbf{Q}:\vec{\nabla}\vec{u})$, with $\mathbf{E} = \tfrac{1}{2} (\vec{\nabla}\vec{u} + (\vec{\nabla}\vec{u})^\top)$ and $\mathbf{\Omega} = \tfrac{1}{2} (\vec{\nabla}\vec{u} - (\vec{\nabla}\vec{u})^\top)$ representing the rate of strain and the vorticity tensors respectively, controls how the nematic field responds to gradients in the flow $\vec{u}$ with the flow alignment $\lambda$.
The rotational viscosity $\gamma$ sets the rate of relaxation towards equilibrium.

The flow field $\vec{u}$ obeys the incompressible Navier–Stokes equations
\begin{equation}        
\frac{D\vec u}{Dt} = \nabla\cdot\mathbf{\Pi},\quad \nabla\cdot \vec u  = 0,
\end{equation}
where the total stress is given by $\mathbf{\Pi} = \mathbf{\Pi}_{\text{viscous}} + \mathbf{\Pi}_{\text{passive}} + \mathbf{\Pi}_{\text{active}}$.
The viscous stress is $\mathbf{\Pi}_\text{viscous} = 2\eta\mathbf{E}$, and the passive stress combines pressure and elastic contributions,
\begin{align*}
\Pi^\text{passive}_{ij} = &-p\delta_{ij} + 2\lambda(Q_{ij}+\delta_{ij}/2)(Q_{lk}H_{kl}) \\
-& \lambda H_{ik}(Q_{kj}+\delta_{kj}/2) 
- \lambda (Q_{ik}+\delta_{ik}/2)H_{kj} \\
&- \partial_i Q_{kl}\frac{\delta \mathcal{F}}{\delta\partial_j Q_{lk}} 
+ Q_{ik}H_{kj} - H_{ik}Q_{kj}.
\end{align*}
The active stress $\mathbf{\Pi}_{\text{active}} = -\zeta \mathbf{Q}$ injects energy into the system through the orientational order.
The activity parameter $\zeta$ therefore tunes the strength of non-equilibrium driving.

\subsection{Fluctuating active nematics} \label{sec:FN}
To separate the effects of explicit activity from those of stochastic forcing, we use a fluctuating nematic model introduced in~\cite{bonn2022fluctuation}.
This model modifies the standard active nematic equations by replacing the deterministic active stress with stochastic noise, resulting in the following equations:
\begin{align}
\frac{D \mathbf Q}{D t} - \mathbf{S}&= -\frac{1}{\gamma}\mathbf H + \boldsymbol{\xi}^Q,\\
\frac{D \vec u }{D t} &= \nabla\cdot \mathbf \Pi + \nabla \cdot \boldsymbol{\xi}^u.\label{eq:fluct}
\end{align}
Again, $\mathbf{H}$ is the molecular field and $\gamma$ is the rotational viscosity, while the stress $\mathbf{\Pi}$ remains the same as in the active nematic model but without the active term.
The noise terms $\boldsymbol{\xi}^Q$ and $\boldsymbol{\xi}^u$ are Gaussian and have zero mean.
Their variances are given by
\begin{align}
\langle \xi^Q_{ij}(\vec x,t) \xi^Q_{kl}(\vec x',t') \rangle &= \frac{2}{\gamma}k_B T_Q \mathcal{J}_{ijkl} \delta(\vec x - \vec x') \delta(t - t'), \\
\langle \xi^u_{ij}(\vec x,t) \xi^u_{kl}(\vec x',t') \rangle &= 2k_B T_u \eta \mathcal{J}_{ijkl} \delta(\vec x - \vec x') \delta(t - t'),
\end{align}
where $k_B$ is the Boltzmann constant and $\eta$ is the solvent viscosity.
The tensor $\mathcal{J}_{ijkl} = \delta_{ik}\delta_{jl}+\delta_{il}\delta_{jk} - \delta_{ij}\delta_{kl}$ ensures that fluctuations preserve the symmetry and tracelessness of $\mathbf{Q}$ and the symmetry of the stress tensor.

The strengths of nematic and fluid fluctuations are controlled by $T_Q$ and $T_u$ respectively.
As there is no active term involving $\zeta$, the model is explicitly passive when $T_Q = T_u$, because detailed balance is preserved.
However, detailed balance is broken when $T_Q \neq T_u$, allowing the system to behave as an effectively active fluid driven by stochastic forcing.
This framework provides a clean way to test whether conformal invariance arises from general non-equilibrium fluctuations rather than specific microscopic mechanisms.
In our analysis, we compare the geometric statistics of zero vorticity contour lines for both fluctuating passive ($T_Q = T_u$) and fluctuating active ($T_Q \neq T_u$) regimes.

\subsection{Computational implementation of nematic models} \label{sec:computational}
All simulations of active fluid models are performed using a hybrid lattice Boltzmann approach~\cite{thampi_active_2016}, with the implementation of fluctuations following~\cite{bonn2022fluctuation, adhikari_fluctuating_2005}.

The simulations are carried out in a periodic square domain of side length $L = 2048$.
The system is initialised with a small amount of noise $n_0$ in the nematic order parameter field to avoid metastable states.
Each system is first equilibrated until the number of defects reaches a steady state, where applicable.

The parameters listed in Table~\ref{tab:parameters} correspond to a low Reynolds number regime with {$Re \sim O(10^{-1})$}.

\begin{table}[ht]
\centering
\caption{Parameters used for hybrid lattice Boltzmann simulations of active and fluctuating nematics, and their dimensions in length $L$, mass $M$, and time $T$.}
\bigskip
\label{tab:parameters}
\begin{tabular}{l|l|l|l}
\hline
Parameter            & Symbol    & Value & Dimension \\ \hline \hline
Bulk constant        & $C$       & 1     & $M/T^2$   \\ \hline
Elastic constant     & $K$       & 0.05  & $ML^2/T^2$\\ \hline
Flow alignment       & $\lambda$ & 1     & 1         \\ \hline
Rotational viscosity & $\gamma$  & 20    & $M/T$     \\ \hline
Solvent viscosity    & $\eta$    & 100/6 & $M/T$\\ \hline
LB relaxation time   & $\tau_{LB}$& 1    & $T$       \\ \hline
Square domain edge length & $L_x$& 2048  & $L$       \\ \hline
Initial noise in alignment& $n_0$& 0.05  & 1         \\ \hline
Numerical integration time& $\Delta t$& 1& $T$       \\ \hline
\\ \hline
Activity             & $\zeta$   & [0, 0.1]&$M/T^2$  \\ \hline
Flow fluctuations    & $k_B T_u$     & [0, 0.1]&$ML^2/T^2$\\ \hline
Nematic fluctuations & $k_B T_Q$     & 0.1    & $ML^2/T^2$\\ \hline
\end{tabular}
\end{table}

\begin{figure}
    \centering
    \includegraphics[width=1\linewidth]{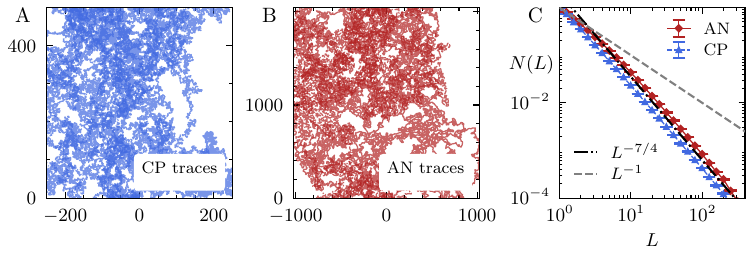}
    \caption{\rev{\textbf{Comparison of traces between critical percolation (CP) and active nematics (AN).}
    (A) Sample of $3$ spanning cluster boundaries for critical site percolation.
    (B) Sample of $3$ vorticity nodal lines for a high activity active nematic ($\zeta= 0.03$).
    (C) Yardstick measurement of critical percolation cluster boundaries \rrev{$N=201$} and active nematics nodal lines \rrev{$N=181$} shows a clear fractal dimension of $d_f = 7/4$, with an offset caused by the larger small distance cutoff in the active nematic. \rrev{Data as mean $\pm$ S.E.M.}
    }}
    \label{fig:ED5}
\end{figure}

\rev{\subsection{Comparison of Critical Percolation and Active nematics} \label{sec:cpvsan}
In order to fully show the similarities between the cluster boundaries in critical percolation and the vorticity nodal lines in active nematics, we show examples of both lines and compare them using the yardstick method which measures their fractal dimension.
}
\rev{
We find that in both cases, the traces are visibly scale free (\textbf{Fig.\ref{fig:ED5}A,B}), with the difference that while the critical percolation trace has a small range cut-off set by the lattice spacing, for active nematics that cut-off is set by elasticity and is therefore somewhat longer.
This is also visible in the measure of the fractal dimension (\textbf{Fig.\ref{fig:ED5}C}), where both curves follow $N(L) \sim L^{-7/4}$ but the crossover into that regime is at a higher value of $L$ for the active nematic trace leading to the offset between the curves.
}

\section{Conformal transition in Fluctuating active nematics} \label{sec:fluctnem}

In order to test the necessity of active driving, we use the fluctuating nematic model (\ref{eq:fluct}) where, by tuning the nematic fluctuations and flow fluctuations, $T_Q$ and $T_u$ respectively, we can break detailed balance by setting $T_Q \neq T_u$, or conserve detailed balance when $T_Q = T_u$~\cite{bonn2022fluctuation}.
We therefore test whether the transition to fully developed active turbulence is a genuinely active phenomenon that arises from detailed balance breaking.

For different values of the reduced distance from equilibrium, {$u = (T_Q - T_u)/(T_Q + T_u)$}, such that $u=0$  is at equilibrium and $u=1$ is maximally distant from equilibrium, we first measure whether the systems exhibit scale free contour lines with a fractal dimension $d_f = 7/4$, as expected for $SLE_6$~\cite{beffara_dimension_2008}.
We measure the fractal dimension $d_f$ using the established yardstick method, in which the contour is measured with yardsticks of different lengths $L$ to determine the scaling of the required number of sticks $N(L) \sim L^{-d_f}$~\cite{mandelbrot_how_1967, noseda2024conformal}.

\begin{figure}[ht]
    \centering
    \includegraphics[width=0.8\columnwidth]{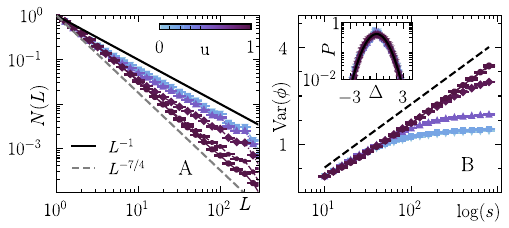}
    \caption{\textbf{Equilibrium fluctuating nematics break $SLE_6$}
    \textbf{(A)} Yardstick measurements for different values of non-equilibrium driving $u$.
    \textbf{(B)} Winding angle measurement, with guide to the eye at $\var(\phi) = 6/7\log(s)+c$.
    Inset shows Gaussian distribution of $\phi$ values.
    \rrev{For (A) and (B), data as mean $\pm$ S.E.M., $N>50$}.
    }
    \label{fig:ED6}
\end{figure}

When the system is at equilibrium, $u = 0$, we find a simple scaling in the nodal lines (zeros of the vorticity field) $N \sim L^{-1}$, corresponding to regular curves. 
As we increase the non-equilibrium driving $u$, the system transitions to a scale free regime with $N \sim L^{-7/4}$, consistent with $SLE_6$ (Fig.~\ref{fig:ED6}A).

To test for conformal invariance, we next measure the winding angle $\phi$, which is known to scale with contour length $s$ as $\var(\phi) = (6/7)\log(s) + c$ for $SLE_6$ curves~\cite{duplantier_winding-angle_1988, boffetta_how_2008}.
We define the winding angle $\phi$ as follows.
Discretising the curve into straight line segments indexed by $i$, each pair of consecutive line segments $i$ and $i+1$ meets at an angle $\alpha_i$, where $\alpha = 0$ if the segments are parallel \rev{(Fig.\ref{fig:ED2}B)}.
The total winding angle of a curve of length $s$ is then calculated as $\phi(s) = \sum_0^s \alpha_i$.
We find that when $u$ is large and the system is active, $\var(\phi)$ follows the expected $SLE_6$ scaling, while at lower $u$ it deviates from this behaviour (Fig.~\ref{fig:ED6}B).
The winding angle of a conformally invariant curve must also be Gaussian distributed~\cite{duplantier_winding-angle_1988}, which we confirm for high $u$ in the inset of Fig.~\ref{fig:ED6}B.

While the left passage probability is inconclusive in this system, likely because domain spanning interfaces are rare and the assumptions of the chordal SLE geometry are not strictly satisfied, the driving function measurements show $SLE_6$ behaviour for active systems (not shown).

We therefore find that for equilibrium fluctuating nematics ($T_Q = T_u$), all measures deviate from $SLE_6$, demonstrating that activity and detailed balance breaking are required to reach the fully developed active turbulence phase.





\section{Geometric and mechanical percolation} \label{sec:vortex_percolation}

\subsection{Detection of vortex centres} \label{sec:vort_detect}
To locate the centres of rotational flow structures, we computed the discrete winding number $m$ of the velocity field $\mathbf{v} = (v_x, v_y)$~\cite{hoffmann2021robustness}.  
Grid points satisfying the topological condition $||m|-1| < \varepsilon$ (with $\varepsilon = 0.3$) were identified as candidate vortex centres.

We then applied a geometric refinement based on the local flow structure.  
For each candidate, the velocity field within a radius of three grid points was decomposed into radial and tangential components.  
A defect was accepted as a vortex centre only if the tangential kinetic energy contributed more than 75\% of the total local kinetic energy ($E_{\theta}/E_{\mathrm{tot}} > 0.75$).  

Weak excitations were filtered out by requiring the vorticity magnitude at the defect centre to exceed the spatial standard deviation of the vorticity field,
\begin{equation}
|\omega| > \sigma_{\omega}.
\end{equation}
The sign of $\omega$ indicates the rotation direction vortex ($+1$ for clockwise rotation and $-1$ for counter-clockwise).  
Detections within three grid spacings were merged to yield a single vortex centre per coherent structure. 
A representative snapshot is shown in Fig.~\ref{fig:ED7}A.

\subsection{Geometric percolation} \label{sec:geom_perc}
To quantify the geometric connectivity among vortex centres, we performed a percolation analysis using random geometric graphs~\cite{pike1974percolation, penrose2003random}.  
For each snapshot, the set of $N$ vortices located at positions $\{\mathbf{x}_i\}$ were treated as nodes in a graph.  
Two vortices $i$ and $j$ were connected if their Euclidean separation was smaller than a probing threshold $r$.  

We computed the percolation order parameter $P_\infty$, defined as the fraction of vortices belonging to the largest connected cluster,
\begin{equation}
    P_\infty(r) = \frac{1}{N} \max_k |s_k(r)|,
    \label{eq:percolation}
\end{equation}
where $|s_k|$ is the size of the $k$th cluster.  
To compare systems at different activity strengths $\zeta$, the probing distance $r$ was normalised by the characteristic length $\ell = r(g_{max})$, extracted from the peak of the pair-correlation function $g(r)$ (see Fig.~\ref{fig:ED7}B).  

\begin{figure}[ht]
    \centering
    \includegraphics[width=0.85\linewidth]{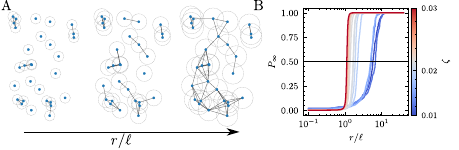}
    \caption{\textbf{Geometric connectivity and percolation of vortices} 
    \textbf{(A)} Schematic illustration of random geometric graph construction.  
    Vortex centres (dots) are connected if their distance is smaller than the probing radius $r$, which increases relative to the characteristic length $\ell$ from left to right.  
    Dashed circles indicate the effective interaction radius, equal to half of the connection threshold.  
    \textbf{(B)} Spanning-cluster fraction $P_{\infty}$ as a function of normalised radius $r/\ell$ for various activity levels $\zeta$ (color scale).  
    Higher activity leads to earlier onset of percolation at smaller scales.  
    The horizontal line marks $P_{\infty}=0.5$.  
    The characteristic length $\ell$ corresponds to the first peak of $g(r)$. For each line, $N=25$.}
    \label{fig:ED8}
\end{figure}

By gradually increasing $r/\ell$, we tracked the growth of the largest connected component (schematic in Fig.~\ref{fig:ED8}A) and identified the effective percolation threshold $r_c$ as the point where $P_\infty = 0.5$ (Fig.~\ref{fig:ED8}B).

\subsection{Mechanical percolation} \label{sec:mech_perc}
To characterise the mechanical rigidity of the vortex network, we constructed a graph where neighbouring vortices were connected according to the $\alpha$-shape complex~\cite{Edelsbrunner1994Three-dimensionalShapes, edelsbrunner1993union} with parameter $\alpha = 2\ell$.  
Rigidity was then analysed using the Pebble Game algorithm~\cite{jacobs_generic_1995} following Petridou \textit{et al.}~\cite{petridou2021rigidity}.  
This graph-theoretic method decomposes the network into rigid and floppy regions by testing whether each added bond introduces an independent constraint or a redundant one, thereby identifying maximally rigid clusters~\cite{jacobs_generic_1995, jacobs1997algorithm}.  

For each configuration, we calculated the fraction of all vortices belonging to the largest rigid cluster, $f_\mathrm{LR}$, which serves as the order parameter for rigidity percolation.

\section{Persistent homology} \label{sec:PH}

\begin{figure}
    \centering
    \includegraphics[width=1\linewidth]{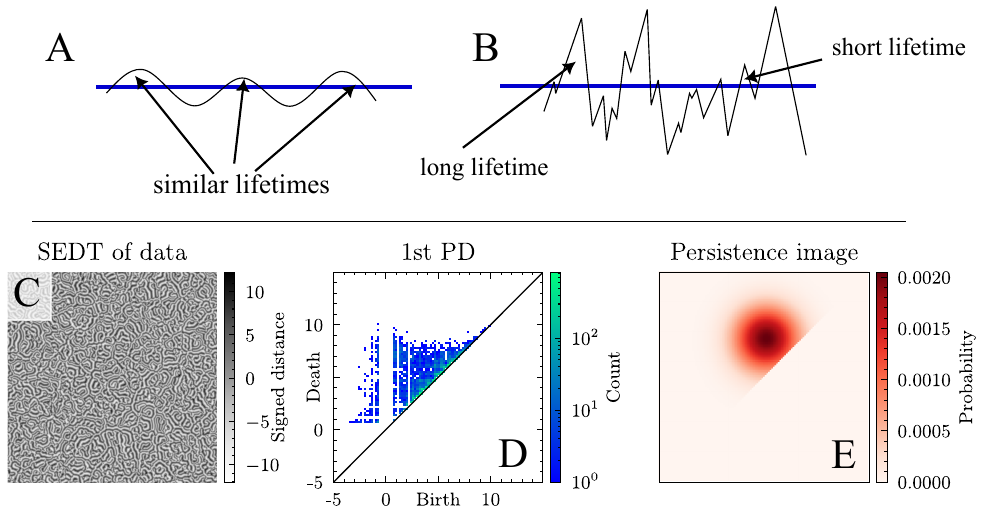}
    \caption{
    \rev{
    \textbf{Upper: Schematic of the persistence homology sublevel set method which essentially corresponds to flooding the landscape of interest.}
    \textbf{(A)} shows that for a regular landscape, birth and death times are similar, resulting in uniform lifetimes.
    \textbf{(B)} shows a more jagged, rough landscape and as a result the birth and death times are more widely distributed, leading also to a wide distribution of lifetimes.}
    \textbf{Lower: Persistence homology workflow for active nematic vorticity field at transition activity $\zeta=0.0215$.}
    \textbf{(C)} Signed Euclidean distance transformed (SEDT) field, showing the (signed) distance from zero vorticity $\omega=0$.
    \textbf{(D)} The first persistence diagram computed from a sublevel set filtration of the data in \textbf{(C)} \rev{shows all birth and death events providing a succinct visualisation of the persistent homology results. }
    \textbf{(E)} Persistence image after~\cite{Adams2015} for the first persistence diagram at (in \textbf{(D)}) \rev{to capture the main features of the persistence diagram.}
    }
    \label{fig:ED9}
\end{figure}

The inputs to our persistent homology calculations are the vorticity field, $\vec{\omega} = \vec{\nabla} \times \vec{u}$.
From these, we compute the signed Euclidean distance transformed (SEDT) field, which encodes how far a given point is from a domain the sign of the vorticity of which is opposite (or equivalently, the zero-vorticity contour line). 
This transformation is sketched in \textbf{Fig.~\ref{fig:ED9}C}. The computed distance is (arbitrarily assigned to be) negative for points inside the domains with negative vorticity.
We normalise these distances by the active length scale, which we estimate to be $\sqrt{K\zeta^{-1}}$. We refer to Table~\ref{tab:parameters} for details on these quantities.

Using standard tools in topological data analysis and persistent homology~\cite{Verri1993, Robins1999, Edelsbrunner2002}, we compute \emph{persistence diagrams} from these distance-transformed fields using the sublevel set method.
In effect we flood the SEDT field and measure how the topology changes during this process, noting birth and death of features as the flood level increases \rev{(See \textbf{Fig.\ref{fig:ED9}}).}

We note that due to vorticity being a pseudo-vector, the choice between the super- and sublevel methods is (also) arbitrary.
Persistence diagrams are then a way of showing the results of the filtration.
An example of the  first persistence diagram is shown in \textbf{Fig.~\ref{fig:ED9}D}. 
In these diagrams, a topological feature is denoted by its birth and death ($b, d$) which are the values the sublevel set for which the feature appears and disappears. 
\rev{
Thus, for the regular landscape in \textbf{Fig.\ref{fig:ED9}A}, the persistence diagram would have a very narrow distribution, up to point-like for a sine landscape, while for the rougher landscape in \textbf{Fig.\ref{fig:ED9}B}, the persistence diagram will be much more widely distributed.}
The zeroth persistence diagram captures the emergence and disappearance of \emph{connected components} in the sublevel sets of our field; and thus each topological feature corresponds to a minimum in the field (and the birth is the value of the field in that minimum), and the death of such a topological feature is the value for which the component ``merges'' with another. For the first persistence diagram, we encode the behaviour of the closed \emph{loops} that emerge during the sublevel set filtration. 
In the first diagram, the birth is the first sublevel set for which a given loop is found, and its death is the value for which the loop closes or merges with another. 

\begin{figure}[!htb]
    \centering
    \includegraphics[width=0.7\linewidth]{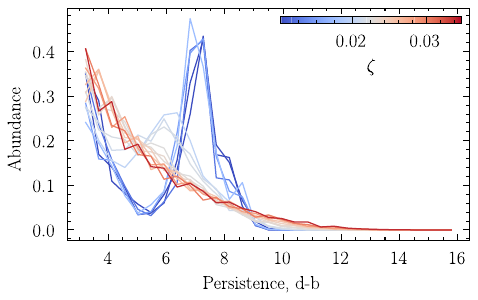}
    \caption{\textbf{Distribution of persistence of the topological 1-features.} 
    \rev{We can see that at the transition, the persistence (lifetime) distribution changes dramatically.}
    }
    \label{fig:ED10}
\end{figure}

We visualize the general trend across our simulation by depicting the persistence, $p = d - b$, and plot their distributions in \textbf{Fig.~\ref{fig:ED10}} (for $p > 3$).
Below the critical activity, the simulations stabilize vorticity domains with specific dimensions; as evidenced by the bimodality of the distributions. 
Once $\zeta$ crosses the critical threshold, the distributions in \textbf{Fig.~\ref{fig:ED10}} cease to be bimodal and transition to long-tailed distributions reminiscent of those refined from Gaussian random fields~\cite{Feldbrugge2019}.

To capture this phenomenon in a quantitative manner across the different values of $\zeta$, we convert these persistence diagrams to \emph{persistence images}~\cite{Adams2015} (example in \textbf{Fig.\ref{fig:ED9}E}) using a Gaussian kernel (with width $\sigma = 4$) weighted by the expression $\arctan\left(c(d-b)^n\right)$ (with $c = 10^{-5}$ and $n = 3$) and subject these to principal component analysis in order to extract the general trends in the data across the different levels of activity. 
These parameter choices are the result of a systematic search, and provide the optimal distinction of the states of our system.
\rev{Although the persistence diagrams, which capture the lifetimes of all features, clearly show a difference between low and high activity active nematics, the use of principal component analysis on the persistence images allows us to reduce the dimensionality of the persistent homology results and therefore crystallise the transition activity in this analysis, and these results agree very well with the transition activity measured with the other methods.}
 
\section*{Data Availability Statement}
The Data are available at https://sid.erda.dk/sharelink/bdcfJSizwj .
\section*{Code Availability statement}
The code for simulating active nematics is available at https://gitlab.nbi.ku.dk/active-intelligent-matter/mass-nematic .
The SLE and connectivity analysis code is available at https://gitlab.nbi.ku.dk/active-intelligent-matter/sle\_connectivity\_transition .
\section*{Methods references}

\newpage
\onecolumngrid

\FloatBarrier
\section*{Extended Data Figures}

\begin{figure}[ht]
    \centering
    \includegraphics[width=0.75\linewidth]{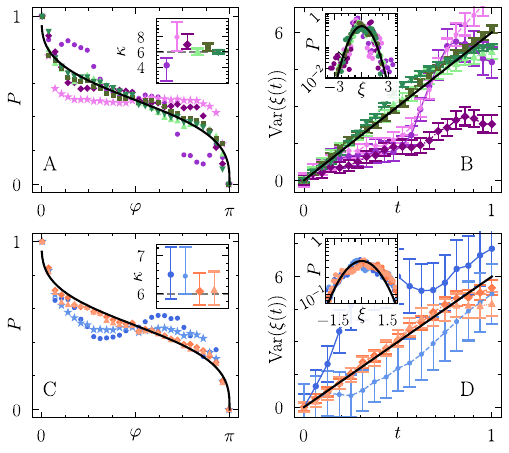}
    \caption{\textbf{Schramm–Loewner evolution analysis of both experimental systems}
    \textbf{(A)} left passage probability of the microtubule–kinesin system at low (purple) and high (green) activity. The inset shows that the high-activity state fits $\kappa=6$.  
    \textbf{(B)} Driving-function analysis for the same system, where high-activity traces follow $\mathrm{Var}[\xi(t)] = 6t$, consistent with $SLE_6$. The inset shows Gaussian statistics of $\xi(t)$ in the high-activity regime.  
    \rrev{For datasets $1-6$, numbered by appearance in (A) inset, the pairs $1\&4$, $2\&5$, $3\&6$ are independent biological replicates, where the smaller number is before activation.
    The number of frames analysed is $N_1 = 29, N_2 = 201, N_3 = 476, N_4 = 174, N_5 = 38, N_6 = 499$.
    }
    \textbf{(C)} left passage probability of the bacterial suspension at low (blue) and high (orange) activity, with the inset confirming $SLE_6$ behavior in the active state.
    \textbf{(D)} Driving-function variance for bacterial suspensions, showing that high-activity states closely follow the $SLE_6$ prediction.
    \rrev{For datasets $a-d$, numbered by appearance in (C) inset, $a-d$ are technical replicates and the number of frames analysed is $N_a = 392, N_b = 924, N_c = 100, N_d = 100$.
    }
    \rrev{For inset $\kappa$ estimations in (A), (C), the errors are as defined in equation \ref{eq:lpperror},
    for driving function measurements the data are presented as mean $\pm$ S.E.M.}
    The data shown in Figure \ref{fig:exp} in the main text is a subset of this data for visibility.
    }
    \label{fig:ED3}
\end{figure}

\begin{figure}[ht]
    \centering
    \includegraphics[width=0.65\linewidth]{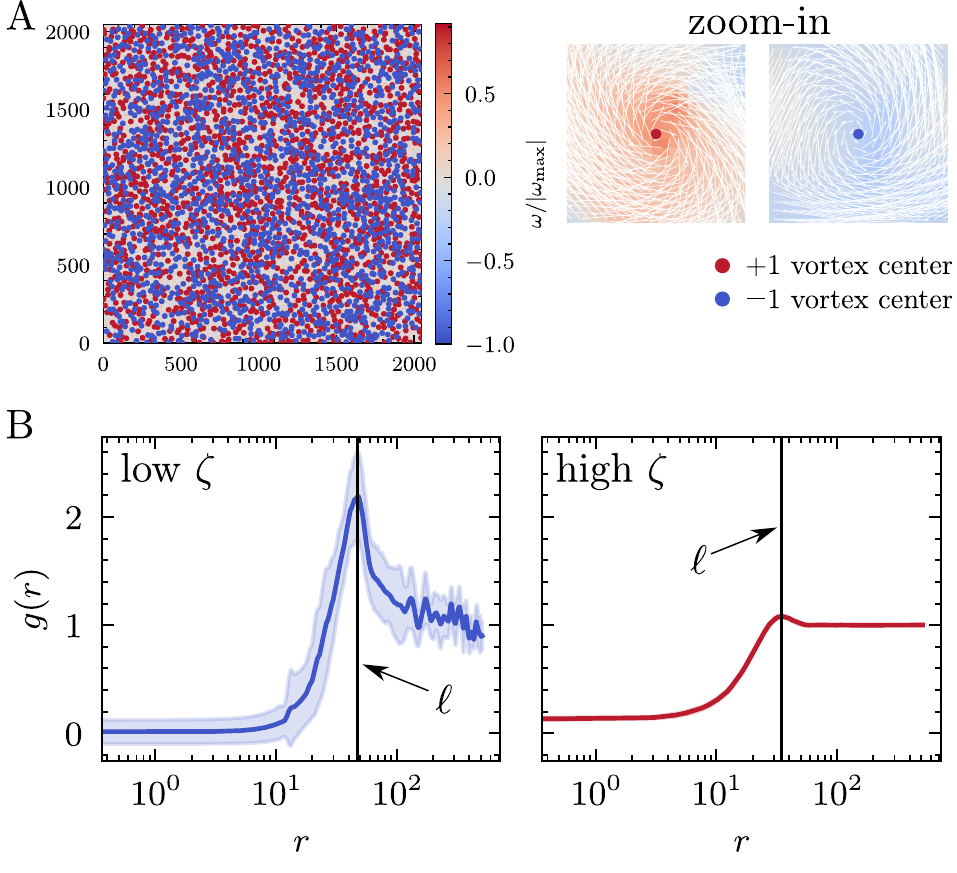}
    \caption{\textbf{Detection and spatial correlations of vortex centres}
    \textbf{(A)} Left: normalised vorticity field $\omega/|\omega_{\max}|$ across the entire domain with all detected $+1$ vortex centres shown as red dots and $-1$ vortex centres shown as blue dots. Right: zoomed-in view of the velocity field (white arrows) overlaid on the normalised vorticity field $\omega/|\omega_{\max}|$ near a single vortex.
    \textbf{(B)} Radial pair-correlation function $g(r)$ of the vortex centres (including both $+1$ and $-1$) for low (blue) and high (red) activity $\zeta$.  
    The vertical line marks the characteristic length scale $\ell$, defined as the position of the first peak of $g(r)$. $N = 25$. 
    Data are shown as mean $\pm$ S.D., whereby at high activity the standard deviation is very small.}
    \label{fig:ED7}
\end{figure}

\end{document}